# SiC-YiG X band quantum sensor for Sensitive Surface Paramagnetic Resonance applied to chemistry, biology, physics.

Jérôme TRIBOLLET

*Institut de Chimie de Strasbourg, Strasbourg University, UMR 7177 (CNRS-UDS),*
*4 rue Blaise Pascal, CS 90032, F-67081 Strasbourg Cedex, France*
E-mail : *tribollet@unistra.fr*

**ABSTRACT**

Here I present the SiC-YiG Quantum Sensor, allowing electron paramagnetic resonance (EPR) studies of monolayer or few nanometers thick chemical, biological or physical samples located on the sensor surface. It contains two parts, a 4H-SiC substrate with many paramagnetic silicon vacancies ($V_2$) located below its surface, and YIG ferrimagnetic nanostripes. Spins sensing properties are based on optically detected double electron-electron spin resonance under the strong magnetic field gradient of nanostripes. Here I describe fabrication, magnetic, optical and spins sensing properties of this sensor. I show that the target spins sensitivity is at least five orders of magnitude larger than the one of standard X band EPR spectrometer, for which it constitutes, combined with a fiber bundle, a powerful upgrade for sensitive surface EPR. This sensor can determine the target spins planes EPR spectrum, their positions with a nanoscale precision of +/- 1 nm, and their 2D concentration down to $1/(20nm)^2$.

Electron paramagnetic resonance[1] (EPR) investigation of electron spins localized inside, at surfaces, or at interfaces of ultrathin films is highly relevant. In the fields of photovoltaic[2] and photochemistry[3], EPR is useful to study the spins of photo-created electron-hole pairs, their dissociation, and their eventual transport or chemical reaction occurring at some relevant interface. In opto-electronics with 2D semiconductors[4], spins of defects limiting device performance can be identified and quantified by EPR. In magnetic data storage science[5] and in spin-based quantum computing science using molecules[6] grafted, tethered, encapsulated or physisorbed on a solid substrate, it is relevant to study by EPR the magnetic properties of those molecules, always modified by their interaction with the substrate[7]. In solid supported heterogeneous catalysis, it is relevant to study spins involved in catalytic reactions, using EPR[8] and eventually spin trapping methods[9]. In structural biology, it is relevant to study by EPR spin labeled proteins[10,11] introduced in polymer supported or tethered lipid bilayers membranes[12,13]. In the context of the development of new theranostic agents for nanomedicine, it is relevant to study ligand-protein molecular recognition events occurring on surfaces by EPR, using for example, bifunctional spin labels[14]. As various nanotechnologies now allow to produce nanoscale thickness samples, one needs to perform sensitive Surface EPR (S-EPR). However, commercial EPR spectrometers have not enough sensitivity[15] for EPR study of those few monolayers thick ultra-thin films, particularly when target spins are diluted and when samples stacking is not possible.

Home-made EPR experimental setups have been developed recently, in the context of quantum sensors[16-20] and quantum computers, reaching single spin sensitivity by optically[17,18,21], electrically[22] or mechanically[23] detected EPR. Some of them achieved the nanoscale resolution imaging, when combined with magnetic devices moving over surfaces[24,25]. Other recent advances in the field of inductively detected EPR have also considerably improved sensitivity, but at the price of operating home-made microwave devices at unconventional millikelvin temperatures[26]. Thus, clearly, there is today a gap between performances of standard X band EPR spectrometers already used worldwide by most of chemists, biologists and physicists, and the ones of the bests unconventional EPR setups found in just few laboratories worldwide.

Here I present the theory of a new Optically Detected Magnetic Resonance (ODMR) based electron spins Quantum Sensor, allowing to study target electron spins of ultrathin paramagnetic samples located on the sensor surface. It has nanoscale resolution in one dimension, a high sensitivity due to spins ensemble ODMR, and importantly, is designed as an upgrade of standard X band pulsed EPR spectrometers. The design of the magnetic properties of the sensor is inspired from the ones of the hybrid paramagnetic-ferromagnetic quantum computer device[27] I previously proposed. However, here, it is adapted to constraints of standard X band (10 GHz, 0.35 T, 5 mm sample access) pulsed EPR resonators and spectrometers and thus to fiber bundle based ODMR[28,29]. The quantum sensor contains two parts. The first is a 4H-SiC semiconductor substrate containing, just below its surface, isolated negatively charged silicon vacancies ($V_2$) used as quantum coherent ODMR spin probes[20,21,30,31,45]. The second part is an ensemble of ferrimagnetic YIG (Yttrium Iron Garnet) nanostripes[32] having narrow spin wave resonances at X band. A fixed spacer fabricated on edges adjust the relative distance between the two parts. Next, I present the fabrication methodology, magnetic and optical properties, and finally spins sensing properties, based on PELDOR spectroscopy[1,10,11,18,33], of this SiC-YiG quantum sensor.

The quantum sensor device proposed can be obtained by fabricating its two parts separately and then integrating them (fig.1 a, b). As said in introduction, the first part of the quantum sensor is a 4H-SiC semiconductor sample, in which silicon vacancies spin probes[20,21,30,31,45] called $V_2$ are created just below the 4H-SiC surface, and on which the ultra-thin paramagnetic film of interest will have to be deposited, anchored or self assembled (fig.1). This is necessary because the spins sensing principle is related to the many long range dipolar couplings that exist between a given single $V_2$ spin probe and the many neighbor target spins (fig.1c), those couplings affecting the spin coherence time of $V_2$ spins probes and being revealed by PELDOR spectroscopy[1,10,11,18,33]. The 4H-SiC sample can be a 4H-SiC substrate terminated on one side by an isotopically purified 4H-SiC grown layer, having no nuclear spins[21] and a very low residual n type doping (< $10^{14}$ $cm^{-3}$) [21]. However, a commercially available 4H-SiC substrate with low n doping and a natural low amount of non-zero nuclear spins is also a good starting point.

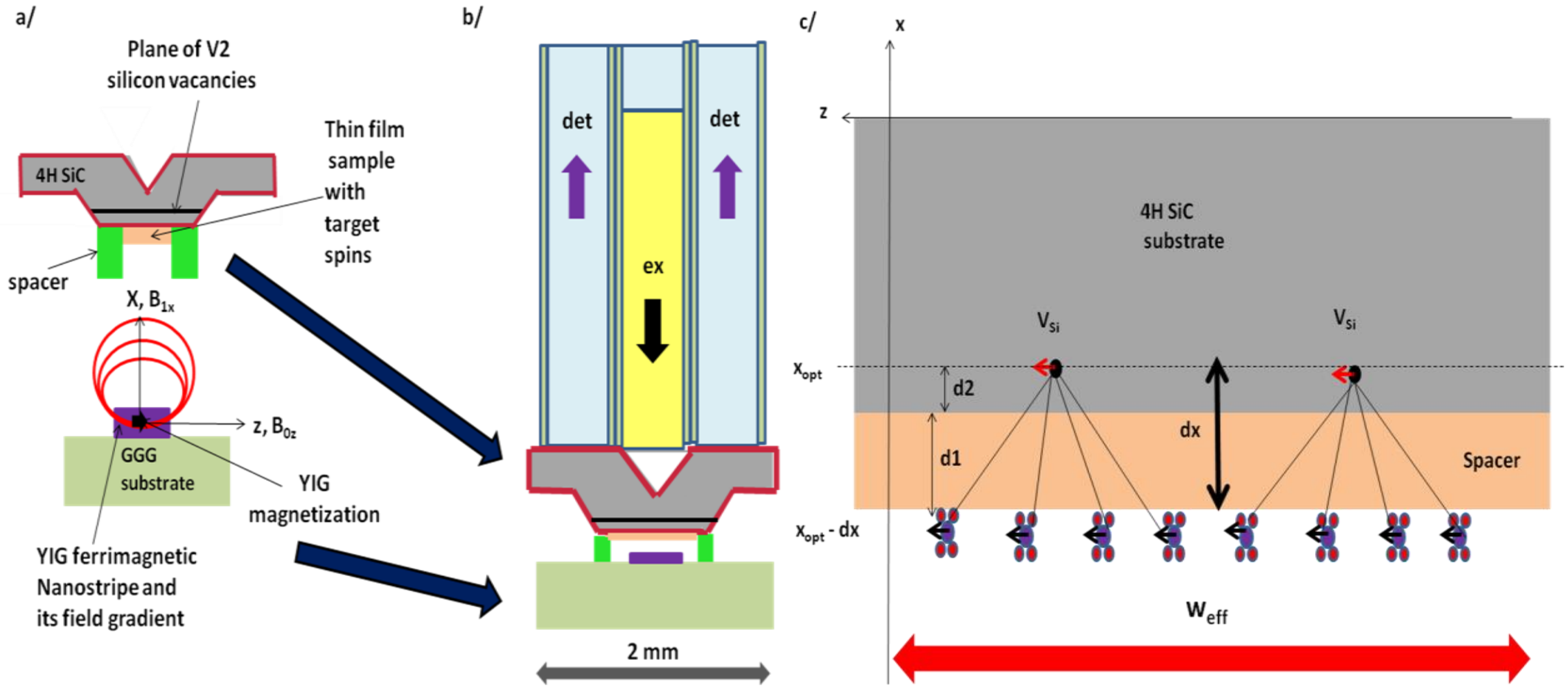

**<u>figure 1</u> : a/ two parts of the Quantum Sensor: the paramagnetic 4HSiC one, with V2 spins on front side of the truncated cone shape island (45°), and a cone shaped dip (45°) on back side; and the ferrimagnetic one, with many identical YIG nanostripes on GGG substrate (only one stripe shown here for clarity, thus not at scale). Also shown on b/, their integration by a spacer (not at scale) and introduction in a standard pulsed EPR spectrometer microwave cavity, as well as the fiber bundle and the GRIN microlens (yellow) used for fiber bundle based ODMR. b/: Zoom showing the many dipolar couplings (dark lines) existing between V2 spins probes in 4H-SiC and target spins in the sample, used for quantum sensing by OD PELDOR spectroscopy. Molecular target spins and V2 probe spins are here separated by a capping layer of few nanometers. $W_{eff}$ indicate the width along z direction over which the dipolar magnetic field produced by a nearby YIG nanostripe can be considered as homogeneous. dx is the distance between the plane of V2 spins and the plane of target molecular spins considered here. d1+d2=dx. Orders of magnitude: $C_{2D,V2}$= 1/(30nm)$^2$ et $C_{2D,Target}$= 1/(5nm)$^2$, dx=10nm, et d2=2nm, d1=8nm, weff =60nm for a nearby YIG nanostripe (T=100nm/W=500nm), whose center is located at a distance xopt=150 nm here from the V2 spins plane.**

The fabrication process of silicon vacancies $V_2$ spins probes in 4H-SiC that I propose here is described on top of fig.2. It is based on an implantation-etching approach, combined with SiC sculpting, in order to define the appropriate photonic structure for the optical excitation and detection of $V_2$ spins probes. After cleaning of the 4H-SiC surface, 5 nm of sacrificial $SiO_2$ are fabricated on the surface of the 4H-SiC substrate (thickness of 400 µm). Those 5 nm of $SiO_2$ can be obtained, either by slow oxidation of the 4H-SiC surface[34] into $SiO_2$ at around 1150 °C, or by a lower temperature thin film deposition method like by PECVD[35] or atomic layer deposition (ALD)[36] at 150°C. High temperature oxidation should advantageously remove residual $V_2$ silicon vacancies initially present in the 3D bulk of the 4H-SiC sample, as $V_2$ vacancies are annealed out[31] at around 700°C. Then, 20 nm of a stopping sacrificial layer of zinc oxide (ZnO) are deposited on top of $SiO_2$/4H-SiC, by sputtering or by ALD. Then 22 keV $As^+$ ions are implanted in this tri-layer sample at a dose comprised between 1.6 $10^{12}$ $cm^{-2}$ and 1.6 $10^{13}$ $cm^{-2}$. The target dose here is around 8.3 $10^{12}cm^{-2}$, which corresponds, according to SRIM simulations (see SI), to a 2D effective concentration of $As^+$ ions in the first 2 nm of 4H-SiC of $C_{2D, As+} = 1/(32nm)^2$. SRIM simulations also indicate that the concentration of $As^+$ ions rapidly decay with depth in 4H-SiC and is almost zero after the first 10 nm of 4H-SiC. SRIM simulations also indicate that such implantation of $As^+$ ions produce 1.3 silicon vacancy per $As^+$ ion in those first 2 nm of 4H-SiC. One can thus consider that we obtain a 2D effective concentration of silicon vacancies $V_2$ in the first 2 nm of 4H-SiC of $C_{2D, eff, V2} = 1/(32nm)^2$. This concentration rapidly decays to zero in the next few nanometers in 4H-SiC. Then, 4H-SiC micro-sculpting is performed either by diamond machining[37,38], by laser ablation[39], by FIB[40] or by another micromachining method[41]. The aim is to produce, on front side, a truncated cone shape island with $V_2$ spins on top, and on back side, a cone shape dip (cone edge angle of 45° in both cases), both cones sharing the same symmetry axis and having an optical quality surface roughness (fig.2 top). Then, ZnO is etched by HCl, and $SiO_2$ is etched by HF[42]. This leads to a sculpted sample with shallow silicon vacancies created mainly 2 nm below the surface of the 4H-SiC truncated cone shape island. A post implantation-sculpting-etching annealing, at a temperature inferior to 600-700°C, can eventually be performed to remove some unwanted created defects. Then, a treatment passivates the truncated cone shape 4H-SiC island surface, like a H+N plasma treatment[43] at 400°C, reducing its surface density of state to 6 $10^{10}$ $cm^{-2}$. Then, eventually (not shown on fig.2), a few nm capping layer, easy to functionalize, can be deposited on this passivated 4H-SiC surface, for example using ALD of silicon oxide at low temperature[36]. Then, a spacer of appropriate thickness, 200 nm here, for example a ring shape spacer made of silicon oxide, is fabricated by standard lithography and deposition, on the edges of the top surface of the 4H-SiC or 4H-SiC/$SiO_2$ island, under which the $V_2$ spins probes were created. The diameter of this top 4H-SiC island surface is around 900 µm. The spacer will allow the integration of the two parts of the quantum sensor device by contacting them (fig.1b). Finally, the few monolayers paramagnetic film of interest can be created on top of the sensor surface. It is either chemically anchored or physically adsorbed on the sensor surface, eventually pre-functionalized. Note also that it is possible to first deposit a nanoscale thickness solid thin film on the sensor surface and then to fabricate a spacer on it, with the appropriate thickness.

The fabrication process of the YIG ferrimagnetic nanostripes array on the GGG (Gadolinium Gallium Garnet) substrate, necessary for the second part of this quantum sensor (fig.2 bottom), follows processes recently published[32,44]. Those processes were successful in producing YIG nanostructured thin films with narrow spin wave resonances at X band[32,44]. Shortly, those processes use a reactive magnetron sputtering system operating at room temperature with a YIG target. The deposition has to be done through a mask fabricated on GGG, obtained by electron beam lithography (fig.2 bottom). After the YIG deposition and mask removal, a thermal treatment at around 750-800°C under air flow or oxygen atmosphere, during around 1 or 2 hours, has to be performed[32,44].

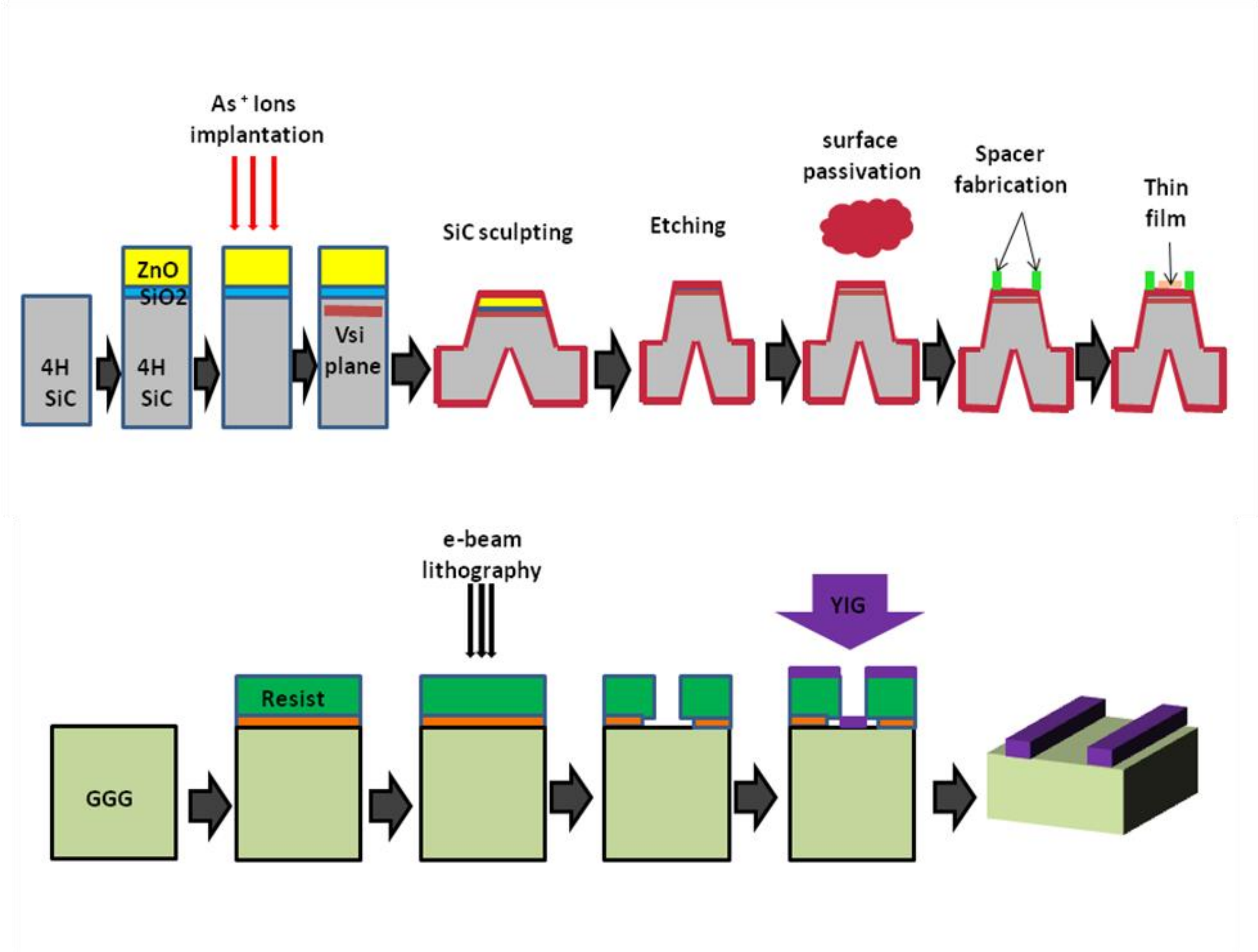

**figure 2:** **Fabrication of the quantum sensor device: 4H-SiC part (top) and YIG/GGG part (bottom); see text for details on the various successive fabrication processes. When possible, and if it is advantageous, the order of some processes can be modified, as long as the key targeted quantum sensor properties are conserved.**

Quantum sensing[16-20,24,25] by optically detected[17,18,21] PELDOR spectroscopy[1,10,11,18,33] (fig.5) is only possible if the $V_2$ spins probes created and coherently manipulated at the microwave probe frequency fs are sufficiently quantum coherent intrinsically, that is without any nearby target spin bath, in order to be able to feel the added spin decoherence[1,17,18,24] produced by the spin bath of the sample of study, when it is driven at the microwave pump frequency fp (fig.5). Let us discuss firstly the electron spin coherence time expected for the spin S=3/2 of a 4HSiC silicon vacancy ($V_2$) [21,30,31,45] created by this fabrication process few nanometers below the surface. Nuclear spin bath spectral diffusion[21] is small in 4HSiC which contains very few non-zero nuclear spins, and it can be eliminated by isotopic purification. Bulk electron spin bath spectral diffusion is small in lightly n-doped 4HSiC and can be reduced by chemical purification and doping control[21]. Spin-lattice relaxation should be quite inefficient for $V_2$ spins probes, in view of the very long spin coherence time of 100 µs observed already at room temperature for bulk $V_2$ spins probes[21,30,31]. Spin decoherence induced by the residual paramagnetic states present at the 4H-SiC passivated surface is negligible for most $V_2$ spin probes, due to the low 2D residual defect concentration after passivation[43] ($6.10^{10}$ $cm^{-2}$). Thus, the dominant intrinsic decoherence process for $V_2$ spins probes in this quantum sensor device is expected to be instantaneous diffusion[1] in 2D, occurring among the $V_2$ spins probes having the same resonant magnetic field, at fixed microwave probe frequency and under the strong dipolar magnetic field gradient produced by the YIG nanostripes. Note that YIG is fully saturated at X band because its saturation field[32,44] is Bsat=1700 G and the external B0 field applied for EPR is around 3500 G.

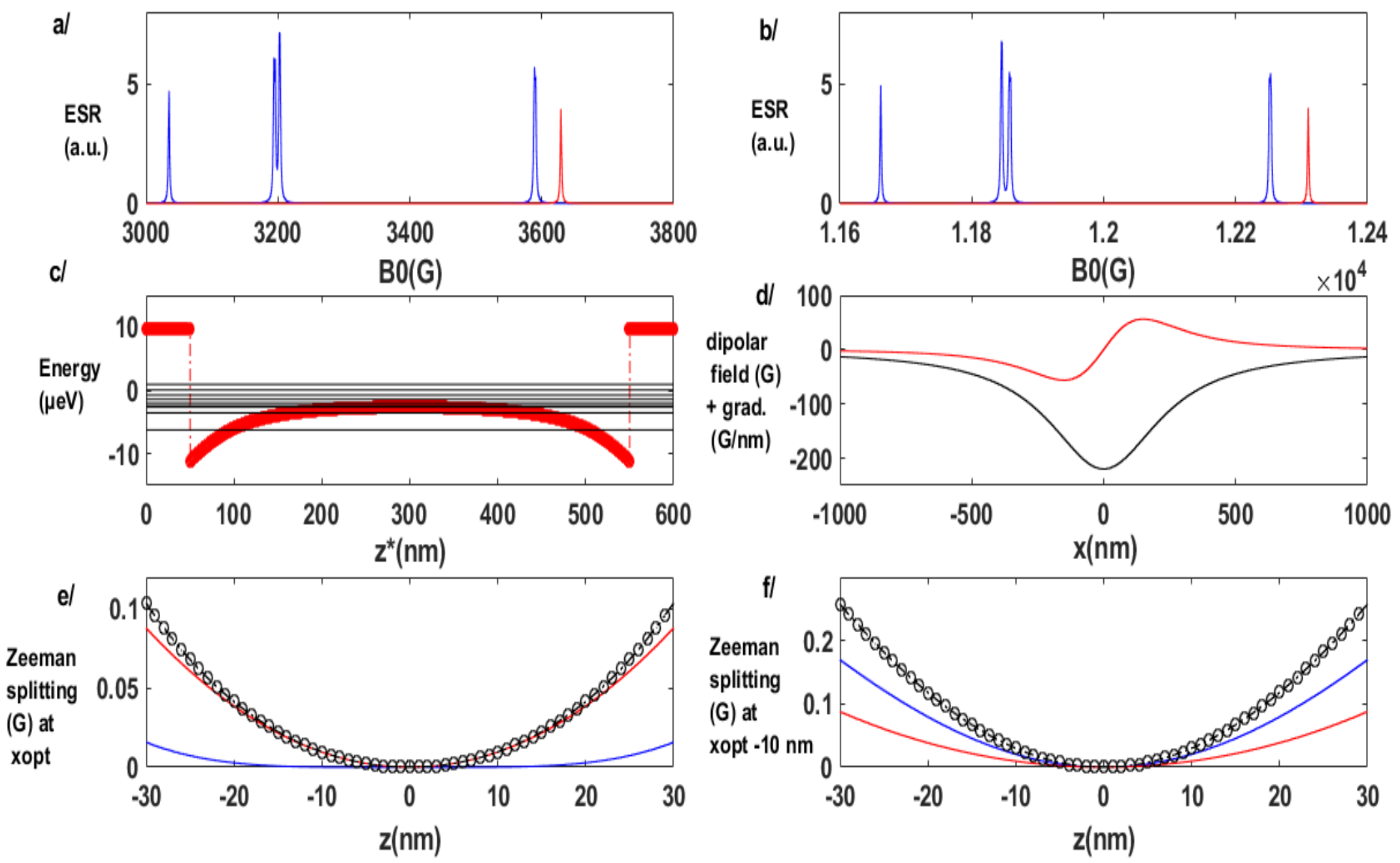

**figure 3 : YIG nanostripes magnetic properties assuming the following dimensions, width W=500 nm , thickness T=100 nm, length L=100µm, and Bsat=1700 G. a/ and b/ Electron spin resonance spectrum at X (9.7 Hz) and Q (34 GHz) band respectively, showing, in blue, the YIG nanostripes spin wave resonances and, in red, the shifted paramagnetic resonance of reference g=2.00 electron spins, placed at $x_{opt}$=150 nm above the YIG nanostripe center (x=0). Paramagnetic and ferrimagnetic resonances have linewidth of 1 G here. c/ One dimensional eigenenergies of the spin waves along z axis (horizontal lines) represented on top of the inhomogeneous effective confining potential inside a YIG nanostripe saturated along its width (here z*=300+z) ; z=0 corresponds to the center of the stripe . d/ z component of the dipolar magnetic field of the YIG nanostripe as a function of x (black), as well as its gradient along x (red) multiplied here by 100 for clarity. e/ and f/ Total effective Zeeman splitting at X band (dot line), expressed in Gauss (thus divided by (g $\mu_B$), assuming g=2.00), as well as its two contributions: the one of Bdz to first order in blue, and the one of Bdx in red to second order, as produced by the YIG nanostripe, respectively at xopt(e/) and at xopt -10 nm (f/), and both plotted versus z, to show the lateral homogeneity of this effective Zeeman splitting.**

The figure 3 summarizes the static and dynamic magnetic properties of the YIG nanostripes. The fig. 3d shows that the maximum magnetic field gradient in the x direction, perpendicular to the GGG and 4HSiC surfaces, is of around 0.5 G/nm and is obtained at a distance $x_{opt}$=150 nm from the center of a given YIG nanostripe. That is why the spacer has to have a thickness of $x_{opt}$ + T/2 =200 nm, such that the $V_2$ spins probes feel the maximum magnetic field gradient. The magnetic field gradient produced by such a YIG nanostripe is not rigorously one dimensional along x. However, as I previously explained in the context of quantum computing[27], locally, around $x_{opt}$ = 150 nm here, and laterally at z=0 +/- 30 nm along z, detailed calculations clearly show (fig. 3e) that in this portion of plane above each YIG nanostripe, the dipolar magnetic field can be considered as laterally homogeneous with a precision of 0.1 G. Even in the portion of plane located at around $x_{opt}$ - 10 nm, and laterally at z=0 +/- 30 nm along z, which is a possible position where target spins could be found, the dipolar magnetic field can be considered as laterally homogeneous with a precision of 0.3 G (fig. 3f). As the $V_2$ spins probes in 4HSiC have a narrow linewidth[21,30,31,45,46] of less than 1 G, with a gradient here of 0.5 G/nm, one can thus consider that all the $V_2$ spins probes located between $x_{opt}$ and $x_{opt}$-2nm (fig 1c), just below the 4HSiC surface, and with z=0 +/- 30 nm along z (weff=60 nm), have the same resonant magnetic field with a precision of around 1 G. As their 2D concentration obtained by fabrication is $1/(32nm)^2$, their decoherence time associated to instantaneous diffusion in 2D is numerically calculated to be $T_{ID,2D}$= 12.5 µs, and is independent of the temperature. Selective microwave pulses[1] can thus excite this $V_2$ spins probes plane, without exciting the other more diluted $V_2$ spins planes located in the next few nanometers of 4HSiC. The $V_2$ plane - target spins plane distance is thus measured here with a precision of around +/- 1nm.

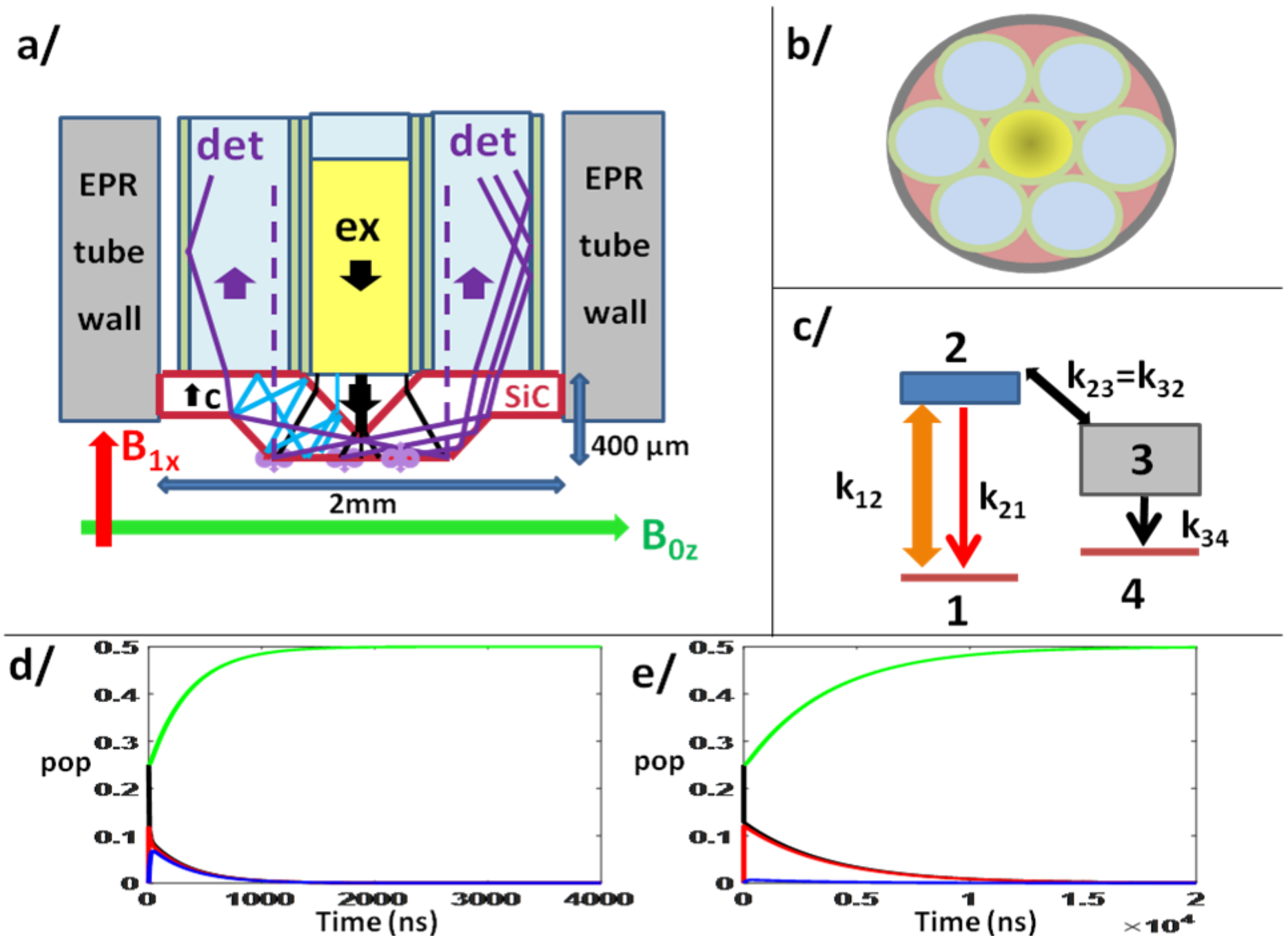

<u>figure 4</u> : Some optical properties of the quantum sensor described here: a/ ODMR setup: fibers bundle (6+1, in blue), GRIN lens (NA=0.5, 0.25 pitch, diameter: 500 µm, in yellow ) for collimation after the central fiber, EPR tube (in gray), and 4H-SiC sculpted sample (edges in red, cone angles are 45°); all are inserted inside a microwave resonator like the MD5 flexline resonator (the YIG part of the sensor, supporting the SiC one, is not shown here for clarity); also shown on a/, static B0z and microwave magnetic field $B_{1x}(t)$, some near surface $V_2$ electric dipoles aligned along the c axis of 4H-SiC (in violet, maximum emission along the z axis , orthogonal to the c axis), and some relevant optical rays for geometric optics investigation of the excitation and collection efficiencies of this new ODMR based setup for quantum sensing. Blue ray is an optical pumping ray with many TIR on SiC faces. Black rays are also optical pumping rays, but TIR are not shown for clarity. Violet rays are photoluminescence rays emitted at 10° with respect to the horizontal and they are still collected by TIR in lateral fibers (NA=0.44, diameter: 500µm). See also zoom in SI. b/ section view of the fiber bundle just above the SiC sample. c/Negatively charged silicon vacancy $V_2$ energy level scheme, explaining the optical readout cycle and the optical pumping cycle. Level names[31,45,46]: 1: (Ground State, S=3/2, $M_{sz}$= -3/2 (or +3/2)), 2: (Excited State) , 3: (Meta-stable excited state) , 4: (Ground State, S=3/2, $M_{sz}$= -1/2 (or +1/2)), $k_{12}$ is laser induced optical absorption/emission rate, $k_{21}$ is photoluminescence rate, $k_{23}=k_{32}=k_{ISC}$ is the intersystem crossing rate, $k_{34}$ is a non-radiative relaxation rate. d/ and e/ : Numerical simulations of populations, based on rate equations, showing the optical pumping[31,45,46] time necessary to saturate the population of $V_2$ spins in the - 1/2 states (green curve) to its maximum value of 0.5 (Note: one can also show that under such OP, the population of $V_2$ spins in the + 1/2 state also saturates to 0.5, using a similar energy level scheme and OP/OD cycles). In d/, k23=k32=1/(17 ns) at 300K[31,45,46], and in e/, k23=k32=1/(1700 ns) assumed at 5K, and for both, $k_{21}$=1/(6ns), $k_{34}$=1/(107 ns), $k_{12sat}$=2.6 $ns^{-1}$. Populations shown: $N_1$ in black, $N_2$ in red, $N_3$ in blue, $N_4$ in green. One finds an optical pumping time of around 20 µs at 5K, and 2 µs at 300K, with those parameters.

It must be also noted here that microwave driving of any spin wave resonance of the YiG nanostripes of the quantum sensor, during the ODPELDOR sequence used for quantum sensing, would add unwanted decoherence[27] to $V_2$ spins probes. That is why the ferrimagnetic insulating YIG nanostripes were carefully designed here such that there is no spectral overlap between their confined spin wave resonances[27] (fig. 3a, b, c), which are narrow in YIG[32,44], and the shifted paramagnetic resonances of the $V_2$ spins probes (fig. 3a, b). Note also that according to my previous theoretical calculation[27], thermal fluctuations of YiG do not contribute to decoherence of $V_2$ spin probes, due to the reduced saturation magnetization of YiG compared to the one of Permalloy previously considered in the context of quantum computing[27]. Note also that, as instantaneous diffusion is temperature independent and as YiG is still ferrimagnetic at room temperature, this hybrid SiC-YiG quantum sensor can be used in principle between 4K and 300K.

The ODMR at X band of the ensemble of $V_2$ spins probes used for sensitive quantum sensing, is based on efficient optical pumping[21,30,31,45,46] (fig. 4 a,c,d,e), as well as on the efficient collection of $V_2$ spins probes photoluminescence[21,30,31,45,46] (fig. 4 a,b), by means of a fiber bundle[28,29], a small GRIN microlens (fig. 1a,b and fig. 4 a,b), and the many total internal reflexion[19] (TIR) occuring both in the sculpted 4HSiC sample (n=2.6) and in the optical fibers (fig 4 a and see also SI). All components of this ODMR setup can be introduced inside standard X band pulsed EPR microwave resonator[1,29] allowing PELDOR spectroscopy, like the MD5 flexline resonator[47], which accept EPR tubes with external diameter up to 5 mm.

One can show that the photoluminescence signal Spl, integrated during T by the photodetector, in the ODPLEDOR sequence (fig. 5a), is given by (see SI): $S_{pl} = S_0.(1\text{-}f)$ , with $S_0 = p_{ex}.p_{coll}.p_{det}.(T/\tau_{V2}).(N_{V2}/8)$ and f, a function that depends on the parameters: $2.t_1$, $2.t_2$, $T_{id,2D}$, td, $C_{2D,T}$, $p_B(f_{pump})$ (see SI for definitions and details). Note that $p_B(f_{pump})$ is equal to 1 when $f_{pump}$ equal the target spins resonant frequency, and 0, when $f_{pump}$ is far off resonance with the target spins resonant frequency. In optimal experimental conditions, the Noise $N_{pl}$ is dominated by optical shot noise, $N_{pl} = (S_{pl}(p_B=0))^{0.5}$. Thus the "net signal" to "noise" ratio R is given by $R=(S_{pl}(p_B=1) - S_{pl}(p_B=0)) / N_{pl}$ . The detailed sensitivity analysis of this quantum sensor (see SI) shows, that in optimal experimental conditions, one could obtain the 200 MHz ODPELDOR spectrum shown on fig. 6b (100 points, one point each 2MHz assumed here) in 1.2 s, with a large signal to noise ratio R=2600.

The numerically simulated (see SI) spins quantum sensing properties, obtained by ODPELDOR (fig.5a), are shown on fig. 6. The figure 6a presents the shifted field sweep EPR spectrum at 9.7 GHz of $V_2$ spins probes located at $x_{opt}$= 150 nm from YIG nanostripes (in green) and of two kinds of target spins S=1 located at $x_{opt}$-dx=145 nm, that is on the sensor surface (in blue and red, see legend for details), as it could be obtained by direct detected EPR, if it would be sensitive enough for Surface Paramagnetic Resonance. The edge spin wave resonance of YiG nanostripes having the highest resonance field at 9.7 GHz has also been added to this spectrum (in pink). The shifted EPR line of $V_2$ at highest field is chosen here for ODPELDOR, which means that $B_{0z}$ is set to this field resonance value, and fs is set to 9.7 GHz, while $f_{pump}$ is scanned during ODPELDOR (fig. 5a). The figure 6b shows the resulting expected X band ODPELDOR spectrum versus $f_{pump}$-fs, scanned over around 200 MHz. The figure 6c indicates how the normalized ODPELDOR net signal to noise ratio (see SI), given by $R/R_{opt}$= 1-$V_{Deer}$(td, dx, $C_{2D,T}$), depends on 1-$V_{Deer}$, $V_{Deer}$ being the DEER[11] signal, and thus how it depends on the relative distance dx between spins probes plane and target spins plane, on the target spin plane concentration $C_{2D,T}$ , and on time constant td. Thus clearly, this SiC-YiG quantum sensor can determine rapidly the target spins plane EPR spectrum and its 2D concentration down to $1/(20nm)^2$, with a sufficiently high net signal to noise ratio, still assuming a $V_2$ spins probes planar concentration of $1/(32nm)^2$, and an associated instantaneous diffusion decoherence time in 2D of $T_{ID,2D}$= 12.5 µs.

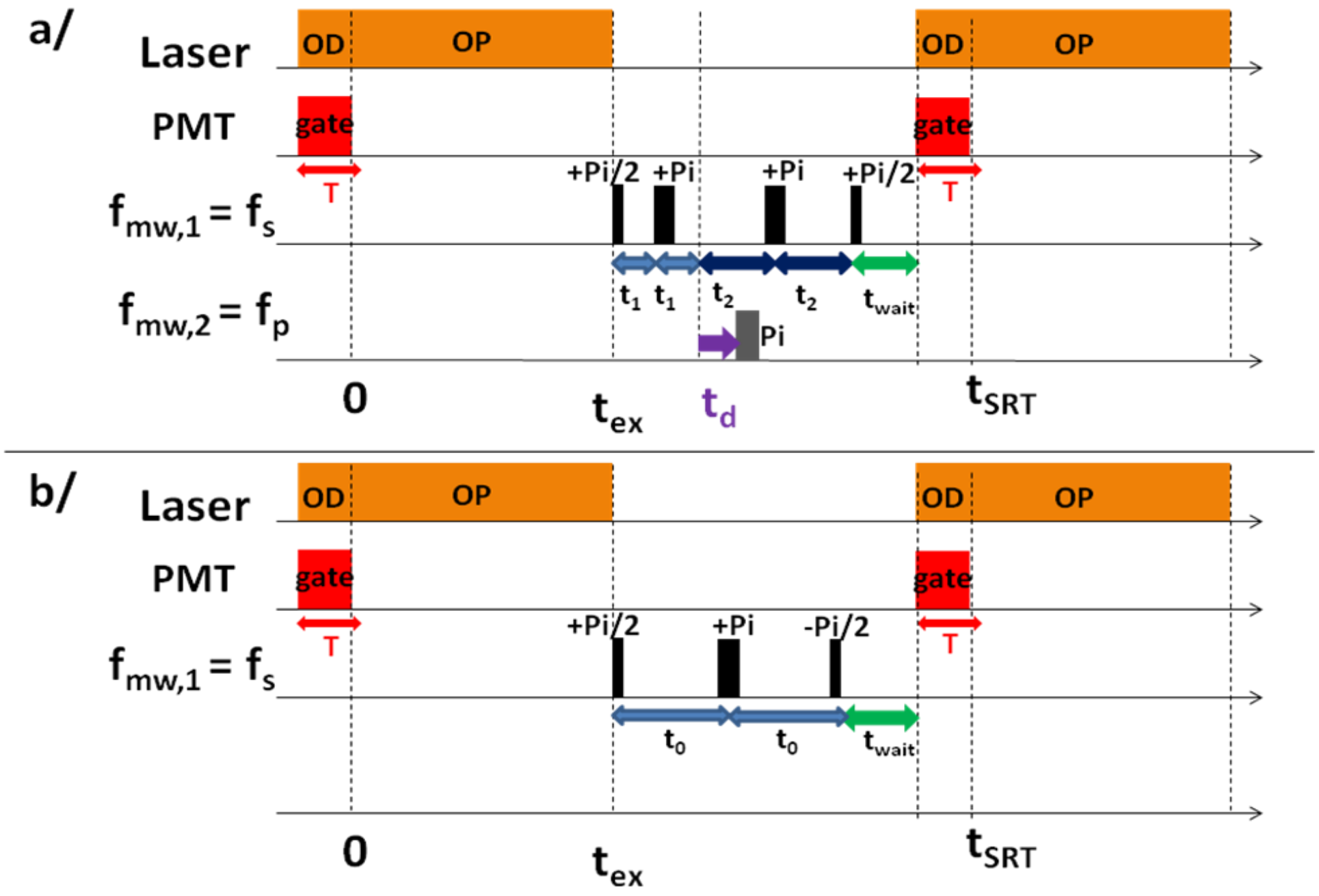

**figure 5: a/ X band OD-PELDOR quantum sensing sequence and b/ X band ODMR spin echo decay sequence for characterization of spin coherence time $T_2$ of $V_2$ spins probes. The spins states -1/2 and +1/2 are prepared simultaneously by optical pumping (laser pulse of 100 µs assumed here). The microwave probe frequency fs, and static field $B_{0z}$, are adjusted to obtain the paramagnetic resonance at this frequency fs with the chosen optically pumped EPR transition of $V_2$ probes spins, either (-3/2 <--> -1/2), or (+1/2 <--> +3/2). Both a/ and b/ time resolved ODMR experiments corresponds nearly to standard PELDOR and Echo Decay experiments[1], but they start after optical pumping and they are complemented by a last +/-Pi/2 pulse in order to transform transverse magnetization Mx ($t_{SRT}$ - T - $t_{wait}$), into populations of $V_2$ spins, which have different spin dependent photoluminescence and relaxation properties under laser excitation. This allows the final optical detection of EPR, the so-called spins ensemble ODMR, by means for example, of a gated Photomultiplier tube (PMT). As a first approximation here, and to better understand the hybrid optical-microwave pulses sequences, spins states -1/2 and +1/2 are assumed Dark states, while spins states -3/2 and +3/2 are assumed Bright photo-luminescent states[45,46].**

Now I compare the sensitivity of this SiC-YiG fiber bundle based ODMR quantum sensor with other setups. Firstly, it must be noted that the same ODPELDOR spectrum as the one of fig.6b could be obtained also in 1.2 s with a quantum sensor having a single $V_2$ spin probe, assuming identical experimental parameters, but at the price of a reduced net signal to noise ratio of only R=2 (see SI). This new spin ensemble quantum sensor[48] is thus 1000 times more sensitive than a similar single spin-based quantum sensor. It is thus advantageous in terms of both measurement time and sensitivity. Of course, ensemble measurements imply an additional statistical averaging of target spins plane properties, which is not present in single spin probe measurements, but such statistics is often a relevant information, like in biology[10,11] and in realistic solid state devices[5,6]. Also, this spins ensemble quantum sensor has a nanoscale spatial resolution in 1D due to the static gradient used, but no scanning and thus no 3D imaging capabilities, contrary to some scanning single spin sensors. Thus, those two kinds of quantum sensors are quite complementary research tools. However, this new quantum sensor based on spins probes ensemble, has not only the advantage of being much more sensitive and faster, but also to be compatible with standard X band pulsed EPR spectrometers, such that it should be widely used in a soon future by many researchers, already using standard EPR and who want to improve its performances. The detailed comparison (see SI) of the sensitivity of standard X band direct inductively detected EPR (DD-EPR) with the one of this quantum sensor upgraded EPR (noted here QUSU-EPR), shows that the sensitivity gain on target spins number is at least of five orders of magnitude. It thus clearly allows to perform surface EPR using this quantum sensor combined with a commercial X band pulsed EPR spectrometer and an optical fiber bundle. This quantum sensor upgraded EPR spectroscopy should thus open new research directions, like in the fields of surface chemistry and photovoltaic, in structural biology and nanomedicine, as well as in optoelectronics, spintronics and quantum information processing.

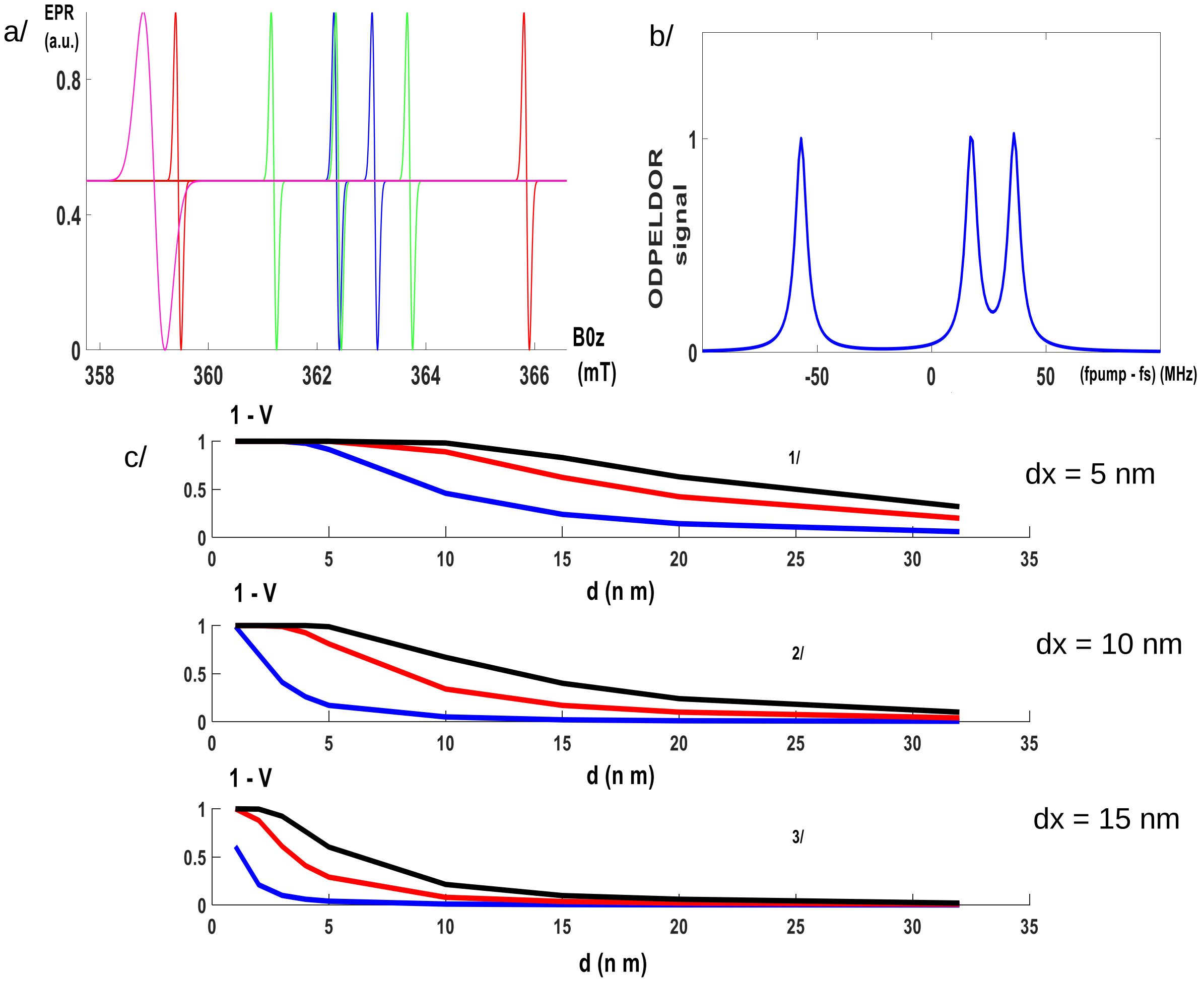

**figure 6 : Spins sensing properties of the quantum sensor. a/ The theoretical shifted field sweep EPR spectrum at fs= 9.7 GHz of spins S=3/2 of V2 spins probes ($g_{iso}$=2.0028, uniaxial magnetic anisotropy along c axis Dc= +35 MHz, $C_{3V}$) located at xopt=150 nm (in green) and of two different ensembles of anisotropic molecular nanomagnets with target spins S=1 ($S_1$=1, $g_{iso,1}$=2.0028,$D_{c,1}$=20MHz, $C_{3V}$, and $S_2$=1, $g_{iso,2}$=2.0028,$D_{c,2}$=180MHz, $C_{3V}$) located at xopt-dx=145 nm here, thus on the sensor surface (assuming 3nm of SiO2 capping layer). $V_2$ spins and nanomagnets are assumed here to have their $C_{3V}$ c axis orthogonal to B0z. EPR simulation in a/ performed with Easyspin software. b/ ODPELDOR spectrum versus fpump-fs, associated to spectrum a/, assuming B0z is set equal to the highest EPR resonance of V2 on a/. c/ Dependence of ODPELDOR normalized net signal to noise ratio (see SI), R/Roptimum=1-V, on the relative distance dx between spins probes plane and target spins plane (dx= 5 nm (1/) , 10 nm (2/), or 15 nm (3/), from top to bottom), as well as on the target spin plane concentration ($C_{2D,Target}=1/(d^2)$, with d in nm). Dark trace is for td= 5µs, red trace is for td=3µs, blue trace is for td= 1µs (see fig.5 for definition of td).**

As a last remark, one can note that this theoretical work, as well as the experimental development[29] of this hybrid SiC-YiG quantum sensor, can be viewed as intermediate steps towards the future development of an intermediate scale hybrid YiG-SiC spins qubits-based quantum computer, following the guidelines I previously published[27]. This not scalable quantum computer design could however still be very useful for efficient quantum simulations of new potential molecular drugs[49]. The advantages of this YiG-SiC quantum computer proposal compared to my previous Permalloy-SiC quantum computer proposal are, the narrow spin wave resonances of YiG, the coherent microwave manipulations of SiC spin qubits at the standard X band, optical initialization and optical detection of EPR of spins qubits ensemble, and probably a high operation temperature for SiC spins qubits, some of them remaining quantum coherent over hundred microseconds, even at room temperature[21,30].

## SUPPLEMENTARY INFORMATIONS

The number of $V_2$ spins probes having the same resonant magnetic field placed at $x_{opt}$=150 nm above a given YIG nanostripe (500 nm*100 nm*100 µm), and within an effective width of $W_{eff}$ =60 nm around z=0 (fig. 1b), is estimated to be at least equal to 3000, taking $C_{2D, V2}$=1/(32nm)$^2$. Assuming the YIG nanostripes are laterally separated by 5 µm, one has an ensemble of around 500 identical YIG nanostripes over the useful square surface of the sensor estimated to be Su=500 µm*500 µm, taking into account the spacer width. Thus, one has around 1.5 $10^6$ identical $V_2$ spins probes on the sensor surface which have the same resonant magnetic field at fixed microwave frequency, that means under the strong gradient produced by the nanostripes. Note also that the surface S* associated to target spins having the same resonant magnetic field is approximately given by S*= (60 nm*100 µm) *500 = 0.3 $10^{-4}$ cm$^2$.

The ODMR at X band of the ensemble of $V_2$ spins probes used for quantum sensing, is based on efficient optical pumping[21,30,31,45,46] (fig. 4a and 4b), as well as, on efficient photoluminescence collection[21,30,31,45,46] (fig. 4b and 4c) of the $V_2$ spins probes in the 4H-SIC sculpted sample, by means of a fiber bundle containing seven fibers (fig. 1a and 4c) and of a small GRIN (gradient index) microlens (fig. 1a), as described in details below.

The central fiber sends exciting light, for example at 780 nm or at 805 nm, along an optical axe common to the GRIN microlens and to the cone shape dip of the 4H-SiC substrate (45° is the half angle of the cone). The GRIN lens, 0.25 pitch plan-plan, allows collimation of the light emerging from the central fiber. Then, by means of a first refraction at the air(Helium)/SiC interface and then by means of the many total internal reflexions (TIR) occuring inside the SiC substrate (n=2.6) (fig.4b), the geometric configuration of the 4H-SIC sculpted sample allows many optical rays to excite the $V_2$ spins located on the useful sensor surface at the top of the truncated cone shape 4H-SIC island. This TIR strategy is inspired from a previous one adopted for sensors fabricated with NV centers in diamond[19], but with here a different sample design, difficult to implement with diamond technology, because diamond is harder than SiC and diamond has not a single defect axis common to all spins probes, like the $V_2$ center in 4H-SiC (the c axis of 4H-SiC is the only axis for $V_2$). This new design allows both optimization of optical excitation and of photoluminescence collection in the restricted volume of an EPR tube of less than 5 mm in external diameter, as required for using standard X band pulsed EPR resonator and spectrometer. Note that the oblique incidence of the exciting light at the sensor surface (incidence angle of around 29° on sensor surface with this design), after the first refraction, provides a non zero optical electric field component parallel to the c axis and thus allows the efficient $V_2$ electric dipole excitation[21,30,31,45,46].Here, I also assume that the optical excitation power at 780 nm or at 805 nm, at the output of the central fiber, is sufficiently high to allow the full saturation of the optical transition, during OD and OP sequences. It was previously shown[46] that the optical power necessary to obtain saturation values of optical $V_2$ spins pumping is inversely proportional to their longitudinal spin-lattice relaxation time $T_1$(T), at the temperature T. As $T_1$(T) increases up to several tens of second at 5K[46], then less than 1 mW at 780 nm spread over a 1mm*1mm square sample is sufficient at 5K for obtaining such optical pumping saturation. Of course, at room temperature, much more power is required, typically more than 100 mw[46]. Thus, from the above considerations, I consider here an optical excitation efficiency for $V_2$ spins located on the useful sensor surface of $p_{ex}$=1.

The photoluminescence of excited negatively charged silicon vacancies $V_2$ in 4H-SiC is emitted at 915 nm at low temperature (zero phonon line[21,30,31,45,46] at 5K). The excited $V_2$ electric dipoles, aligned along the c axis of 4H-SiC, emit their photoluminescence preferentially in the plane perpendicular to the c axis, which means here, at the horizontal. The edges at 45° of the truncated 4H-SiC cone shape island thus allow, by one reflexion, to direct most of the $V_2$ spins probes photoluminescence vertically, towards the six lateral fibers, in which it is efficiently propagated by TIR, till the infrared photoluminescence detector. In order to evaluate more quantitatively the collection efficiency of this fiber bundle based optical setup, defined as the ratio of the collected optical power over the emitted optical power by $V_2$ dipoles, one can use the classical model of a linear dipole aligned along the c axis for the $V_2$ dipole and its emission profile determined using the Pointing vector expression. Using geometric optics (see fig. 4a) and considering the various dimensions of the setup and the relevant refractive index of the materials of the setup ($n_{SiC}$=2.6, $n_{air}$=1, and for fibers $n_{glass}$=1.5 and NA=0.44), one can determine that almost all rays emitted by the $V_2$ dipoles of the useful sensor surface around the horizontal direction at +/- 10° (= π/18 radians), can, after relevant reflexions (TIR) on the 4H-SiC sample surfaces, enter into the lateral optical fibers with a sufficiently small angle such that TIR allows the propagation of those rays without loss till the end of the fibers, towards the photodetector. Considering the Pointing vector expression associated to the $V_2$ dipole in spherical coordinates, one can approximate the collection efficiency $p_{coll}$ by the ratio between the emitted PL and the collected PL, assuming that the PL is collected by the fiber bundle setup when Θ is comprised between (π/2 - π/18) and (π/2 + π/18).

$p_{coll}$ is thus given by the formula:

$p_{coll} = ( \int \sin^3(\Theta)\, d\Theta, \pi/2 - \pi/18, \pi/2 + \pi/18) /( \int \sin^3(\Theta)\, d\Theta, 0, \pi)$

and thus one finds here $p_{coll}$ = 0.25.

The photodetector can be a near infrared sensitive photomultiplier tube with low dark counts, or another low noise infrared photodetection setup. Here I assume a standard infrared photodetector efficiency $p_{det}$=0.01. Note also that the bundle is divided, outside the standard EPR cryostat (like the CF935 from OXFORD for Bruker EPR resonators), into a single fiber, the central one used for optical excitation, and into a bundle of the six lateral fibers collecting the photoluminescence, further directed towards the photodetector.

Now let us evaluate the net signal to noise ratio R of this ODPELDOR experiment and then the sensitivity of this YiG-SiC fiber bundle-based quantum sensor. Starting from the DEER experiment expression[1,11,33], directly related to the ODPELDOR experiment shown on fig. 5a, and considering the optical detection of $V_2$ spins probes and thus the last additional π/2 microwave pulse, one obtains a photoluminescence signal expression Spl, integrated during T by the photodetector, given by: $S_{pl} = S_0.(1-f)$ , with $S_0 = p_{ex}.p_{coll}.p_{det}.(T/\tau_{V2}).(N_{V2}/8)$ and f, a function that depends on the parameters: $2.t_1$, $2.t_2$, $T_{id,2D}$, td, $C_{2D,T}$, $p_B$. The function f is given by $f = \exp(-((2.t_1 + 2.t_2)/ T_{id,2D})^{2/3}).( (1-p_B) + p_B.V_{deer}(td, dx, C_{2D,Target}) )$, where $V_{deer}$ is the standard DEER signal. It can be numerically computed using the linear approximation and shell factorization model[11]. This model was previously introduced for calculating the standard DEER time domain signal in the case of a three-dimensional distributions of spins. Here, this model has been adapted to take into account the bidimensional random distribution of the target spins in their well-defined plane, parallel to the SiC substrate surface. The function $p_B$ depends on the frequency detuning between the microwave pump frequency and the target spin resonance frequency at fixed $B_{0z}$. Thus, $p_B$=1 on resonance, and $p_B$=0 far off resonance for an appropriate duration π microwave pulse. The function $p_B$ is given by the usual probability transition formula describing the Rabi oscillation between the two appropriate spins quantum states under application of a microwave pulse.

In optimal experimental conditions, the Noise $N_{pl}$ is dominated by the optical shot noise, given by $N_{pl} = (S_{pl}(p_B=0))^{0.5}$. Thus the "net signal" to "noise" ratio R is given by the formula $R=(S_{pl}(p_B=1) - S_{pl}(p_B=0)) / N_{pl}$. Thus, introducing $R_{opt}$, the optimal signal to noise ratio, R is given, in the general case, by: $R=R_{opt}*(1- V_{Deer}(td, dx, C_{2D,T}))$, with $R_{opt}$ given by the formula: $R_{opt} = (S_0)^{0.5}. \exp(-((2.t_1 + 2.t_2)/ T_{id,2D})^{2/3}) / (1- \exp(-((2.t_1 + 2.t_2)/ T_{id,2D})^{2/3}) )^{0.5}$. Note here that $R/R_{opt} = 1-V_{Deer}$, that is why $1-V_{Deer}$ is plotted on fig.6. Note also that $R_{opt}$ depends on the spin coherence time $T_{id,2D}$ of $V_2$ spins probes and on the parameters $t_1$ and $t_2$ used in the ODPELDOR experiment. R of course depends on the concentration of target spins $C_{2D,T}$.

Now, assuming a sensor operating with $t_1$=0.5 µs, $t_2$=5.75 µs and $2\,t_1 + 2\,t_2 = T_{id,2D}$ =12.5µs, and assuming $C_{2D,T}=1/(10nm)^2$ , ie sufficiently large such that when td=5 µs, $V_{Deer}(td,C_{2D,T})=0$ ie $1 - V_{Deer}(td, dx, C_{2D,T})=1$ (fig.6c top black curve), then one finds the simple following expression for the best expected signal to noise ratio: $R= (1/e).(S_0)^{0.5}$. With $p_{ex}=1$, $p_{coll}=0.25$, $p_{det}=0.01$, a $V_2$ radiative recombination time $\tau_{V2}$=6ns, and around $N_{V2}=1.5.10^6$ $V_2$ spins probes having the same resonant magnetic field in the sensor (see above), and choosing a photoluminescence integration time per ODPELDOR sequence T=6 µs for example, one finds approximately R=260, for a single "one shot one point" ODPELDOR experiment. The optical re-pumping time of $V_2$ spins is numerically evaluated to $T_{OPump}$= 20 µs at 5K assuming $k_{ISC}$ (5K) =1/ (1700 ns) (see fig. 4e), but the laser pulse is assumed to last 100 µs here for safety, considering the unmeasured value of $k_{ISC}$ at 5K (only known is $k_{ISC}$(300K) =1/17ns at 300K[31], see fig. 4d). The ODPELDOR microwave pulses sequence after optical initialization of $V_2$ spins last around 20 µs, such that the shot repetition time of full ODPELDOR is thus taken here to be $T_{exp}$=120µs. Both $T_{tot,exp}=N_{shot}*T_{exp}$ and $T_{tot}=N_{shot}*T$, increase proportionally to $N_{shot}$, but R only increase proportionally to $(N_{shot})^{0.5}$. Assuming $N_{shot}$=100 per point and a 100 points ODPELDOR spectrum as a function of $f_{pump}$ (1 point each 2 MHz, 200 MHz scanned), one could obtain such a 200 MHz spectrum (see fig. 6b) in 1.2 s with a signal to noise ratio R=2600, assuming negligible hardware and software delays for changing the pumping microwave frequency (otherwise, the experimental time is determined by those delays).

It is here also relevant to compare standard X band direct inductively detected EPR (DD-EPR) sensitivity, with the one of this quantum sensor upgraded EPR method. Assuming a 2D target spins concentration $C_{2D,T}=1/(10nm)^2$, and estimating the surface S* of target spins seen by $V_2$ spin probes and having the same resonant magnetic field to around S*=0.3 $10^{-4}$ cm$^2$ (see above), one finds that around 3. $10^7$ target spins are sensed by the $V_2$ spins probes in 12 ms per point (one point each 2 MHz, 100 shots per point), with R=2600. As in DD-EPR[15] at X band one can typically measure $10^{11}$ spins at 300K or $10^9$ spins at 3K in 1 s with $R_{DDEPR}$ =3 (assuming a 1G linewidth for spins and a 1 Hz detection bandwidth), one finds that in order to obtain R=2600 in 12 ms, one would need $10^{15}$ target spins at 300K or $10^{13}$ target spins at 3K with DD-EPR. The sensitivity gain on target spins number with this quantum sensor is thus comprised between 5 and 8 orders of magnitude. Note that the probe spins sensitivity is considerably higher than the target spins sensitivity, and it could in principle reach the single $V_2$ probe spin sensitivity with long enough accumulation times.

Below, I also provide some results (fig. aux. 1) of the SRIM simulations of 22 keV $As^+$ ions implantation in the trilayer Zn0 (20 nm)/$SiO_2$(5 nm)/4H-SiC (type n <$5.10^{15}$ $cm^{-3}$), allowing, after etching of ZnO and $SiO_2$, to produce shallow silicon vacancies around 2 nm below the surface of 4H-SiC with an average 2D concentration of $C_{2D,V2}$= 1/(32nm)$^2$. SRIM simulations also confirms the advantage of using a trilayer and not just a Zn0/4H-SiC bilayer, because one can see on fig. aux. 2, that some Zn atoms can reach the $SiO_2$ layer due to the implantation process and related collisions ($SiO_2$ is further removed by etching), but not the SiC substrate, thus avoiding pollution with the Zn element of the SiC substrate surface, used for quantum sensing with the silicon vacancies also produced by this implantation process.

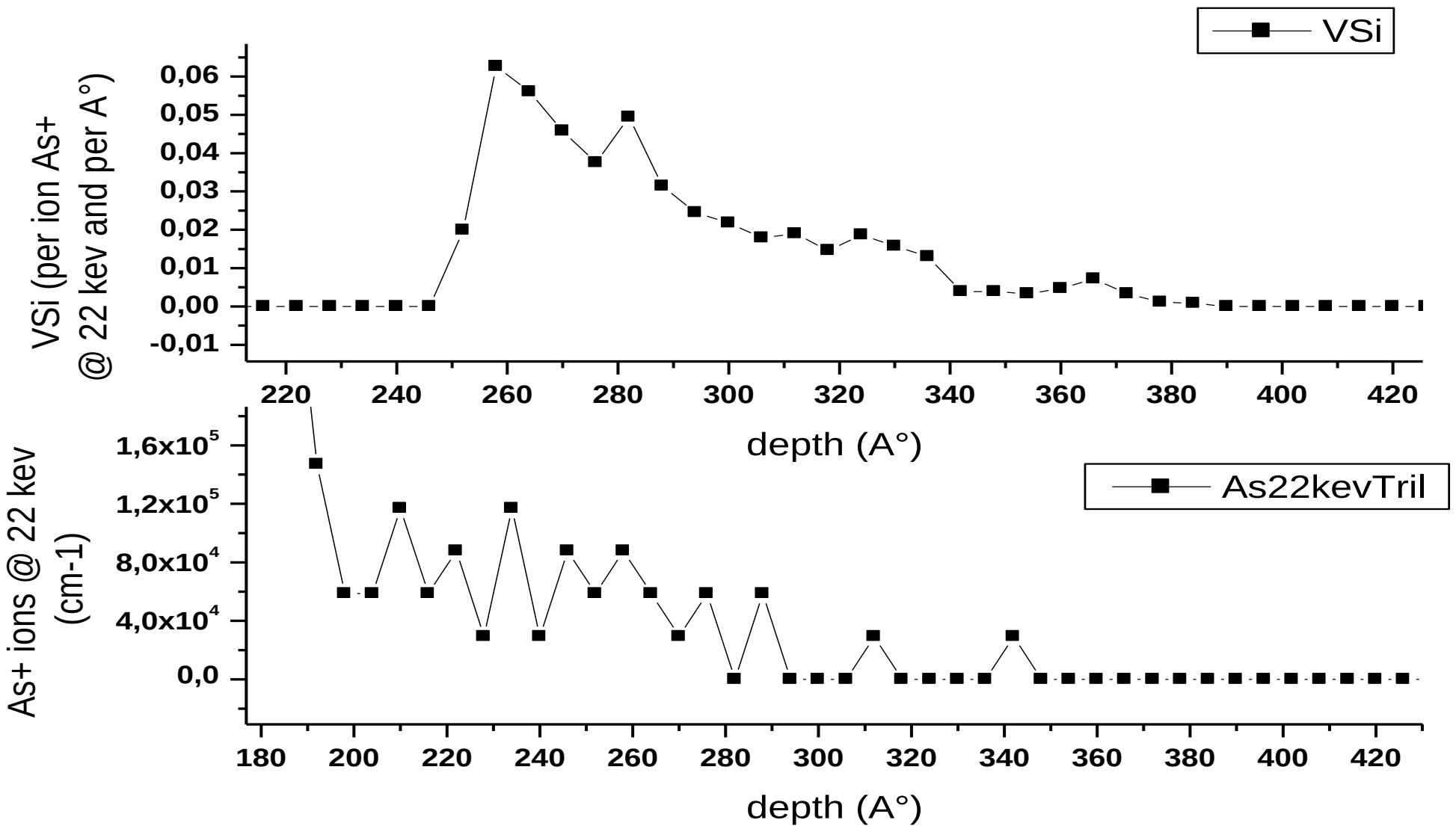

**fig.Aux.1: SRIM simulation of $As^+$ ions implantation at 22 keV in this trilayer system (100 000 shots).**

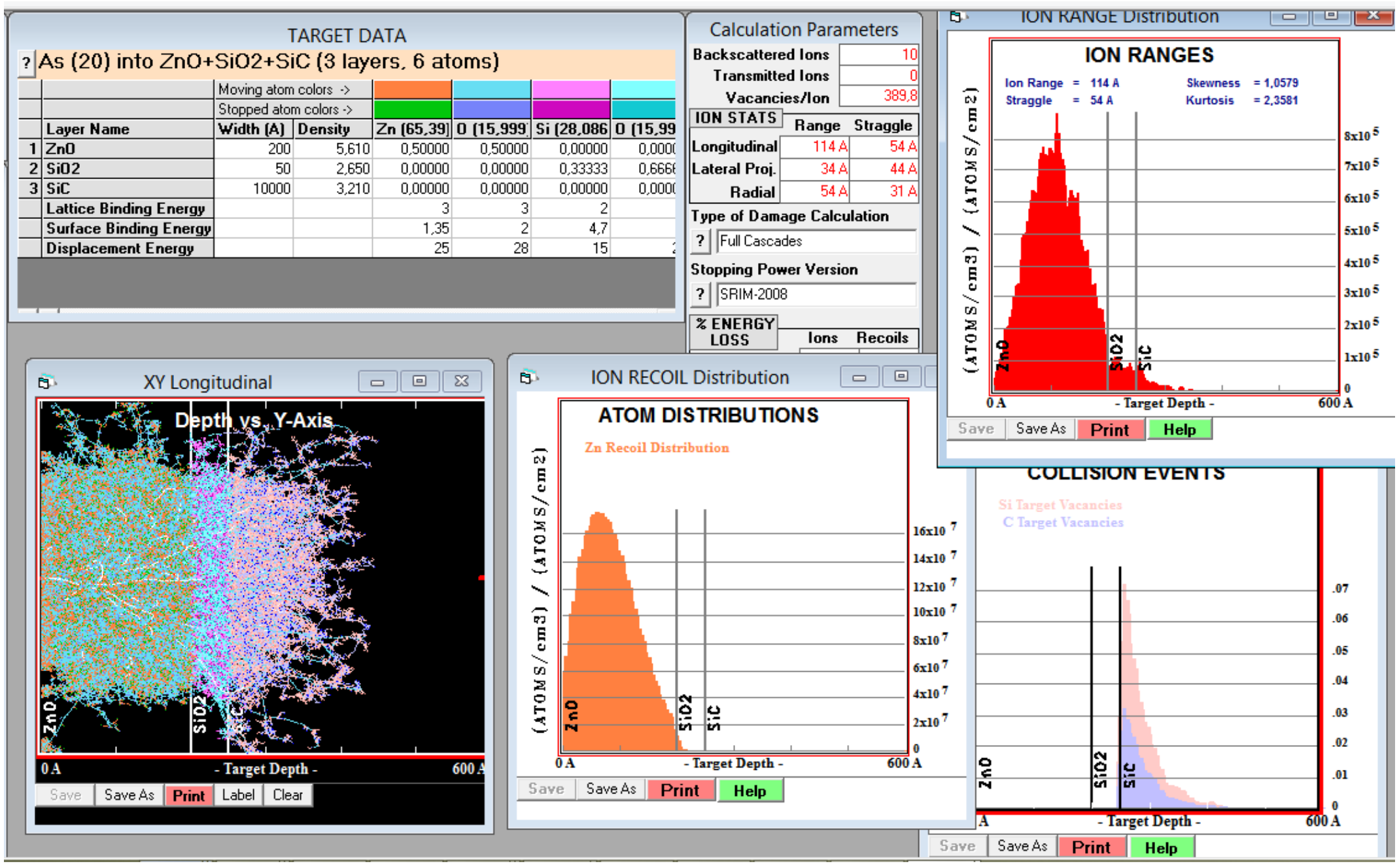

**fig.Aux.2:** SRIM simulation of $As^{+}$ ions implantation at 22 keV in this trilayer system (here 6000 shots).

Below, I also provide (fig. aux. 3) a zoom of fig. 4a used for the discussion of geometric optics in the fiber bundle based ODMR setup adapted to the SiC-YiG quantum sensor described here.

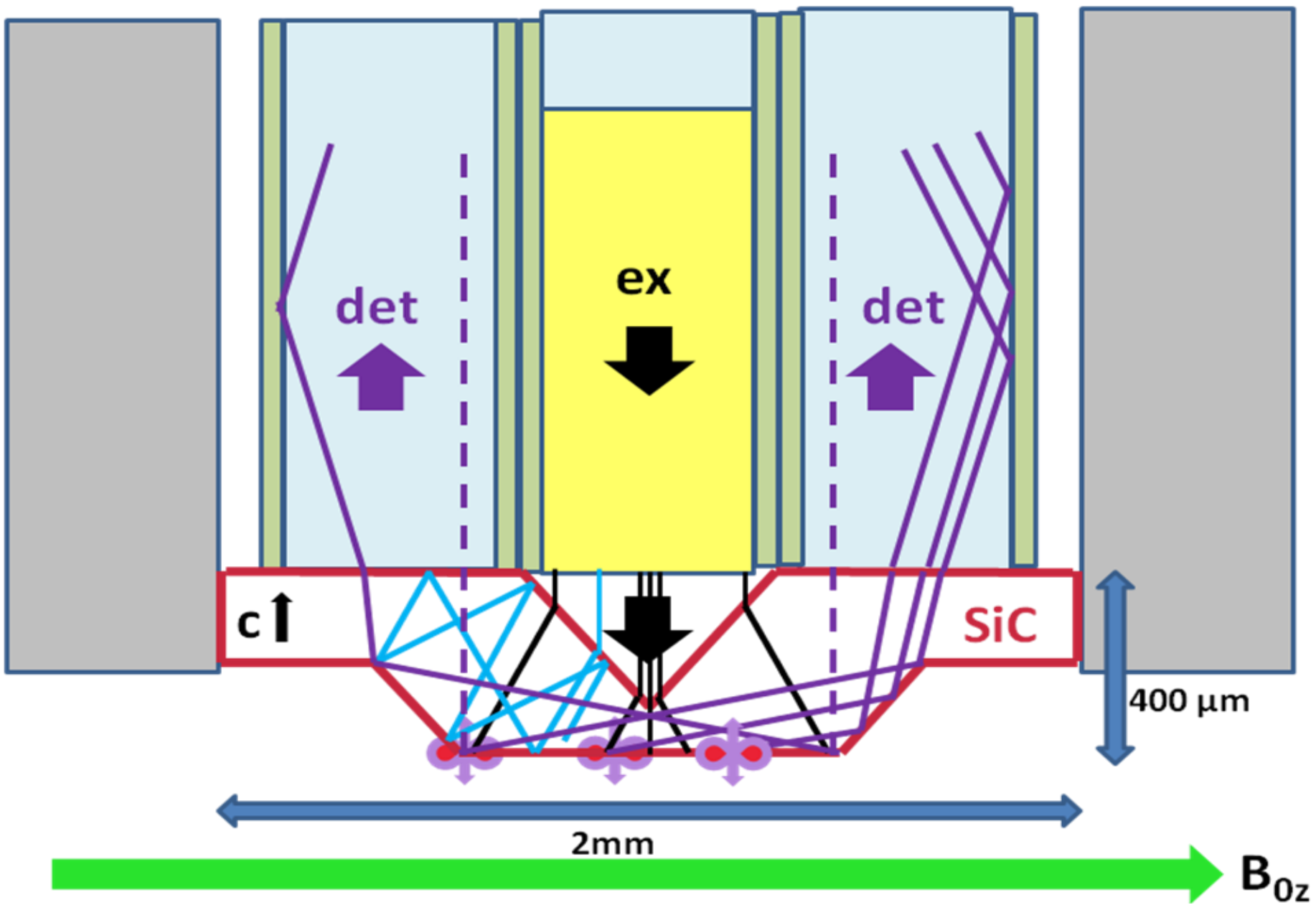

**fig.Aux.3:** Zoom of the setup for geometric optics analysis.